\documentclass[%
 reprint,
 amsmath,amssymb,
 aps,
]{revtex4-2}

\usepackage{amsmath}
\usepackage{graphicx}
\usepackage{dcolumn}
\usepackage{bm}

\usepackage[colorlinks=true, linkcolor=blue]{hyperref}

\begin{document}

\preprint{APS/123-QED}

\title{Quasi-homogeneous geometrothermodynamics of a noncommutative Reissner-Nordstr\"{o}m
black hole}

\author{Alberto Maya}
\email[Correspondence email address: ]{josealbertom66@nu.ac.th}
\affiliation{The Institute for Fundamental Study, Naresuan University,\\ 
Phitsanulok, 65000, Thailand}

\author{Hernando Quevedo}
\email[Correspondence email address: ]{quevedo@nucleares.unam.mx}
\affiliation{Instituto de Ciencias Nucleares, Universidad Nacional Autónoma de México, Mexico City. \\
Dipartimento di Fisica and Icra, Università di Roma “La Sapienza”, Roma, Italy. \\
Al-Farabi Kazakh National University, Al-Farabi av. 71, 050040 Almaty, Kazakhstan.\\ 
}

\date{\today} 

\begin{abstract}
We present the thermodynamic properties of a noncommutative Reissner–Nordström (NCRN) black hole (BH) modeled with Lorentzian distributions. The analysis is carried out using a Legendre-invariant formalism called \textit{Geometrothermodynamics} (GTD) which is applied to a quasi-homogeneous system generated by the NCRN BH. This formalism enables the study of phase transitions by locating Ricci scalar singularities, from which the phase transition points are determined in terms of the thermodynamic variables. We also examine how the noncommutative (NC) parameter $\Theta$ can be interpreted as a thermodynamic variable within quasi-homogeneous thermodynamic laws, highlighting its potential role on phase transitions beyond those well-known characterized by divergences in the heat capacity.
\end{abstract}

\maketitle


\section{\label{sec:level1}INTRODUCTION}

A significant theoretical prediction of General Relativity (GR) is the existence of BHs, whose existence has recently been proved through the direct detection of the BH shadows \cite{Re1}. There are also indications that BHs could be at the center of many galaxies  \cite{Re2}. BHs have been shown to be capable of acting as natural laboratories suitable to test a wide variety of phenomena that challenge the classical common sense and expand the boundaries of physics. 
BHs have gained popularity since the work of Hawking \cite{Re3} and Bekenstein \cite{Re4}, where it was shown that these objects emit radiation and are capable of undergoing phase transitions. As a consequence, there is increasing interest within the physics community in studying BHs as systems with thermodynamic properties,
a research field that is commonly known as black hole thermodynamics \cite{Re5,Re6}.

On the other hand, geometric methods have been intensively applied to study the thermodynamics of BHs from an alternative perspective. This approach can be divided into two different methods called thermodynamic geometry \cite{Re7,Re8} and geometrothermodynamics (GTD) \cite{Re9}. The first one is based upon the use of Hessian metrics in the space of equilibrium states of the corresponding system, whereas GTD uses only metrics that are invariant with respect to Legendre transformations, a property that is derived from standard classical thermodynamics.

As a result of the application of thermodynamic geometry and GTD,  certain geometric properties of the equilibrium space have been shown to represent specific thermodynamic properties of the corresponding system, namely, the Riemann curvature measures the thermodynamic interaction, geodesics represent thermodynamic quasistatic processes, the scalar curvature is associated with the internal microstructure, and
curvature singularities correspond to phase transitions. Some of these concepts were used in GTD to propose a unified geometric description of BHs in Einstein's theory \cite{Re10}.

One of the most important aspects of BH thermodynamics is that it contains information about classical general relativity, statistical physics, and quantum physics. For this reason, it is considered as an essential hint towards a theory of quantum gravity. Nevertheless, although several approaches have been proposed to quantize gravity, a definite theory is still not available \cite{Re11}.

In an attempt to take into account quantum effects in general relativity, noncommutativity (NC) has been proposed as a possible route \cite{Re12, Re13}. In this approach, the spacetime coordinates $x^{\mu}$ are noncommutative objects, i.e., $[x^\mu, x^\nu] = \Theta \varepsilon^{\mu\nu}$ $(\mu,\nu=0,1,2,3)$,  where $\Theta$ is the NC parameter. This approach can be viewed as an analogy of the Heisenberg uncertainty principle, which is well known to represent the quantization of the phase space in quantum mechanics. A notable mathematical consequence of the NC formulation is the change in the classical Dirac delta distributions, which, in a strict derivation using the Moyal product, can be replaced by a Gaussian distribution \cite{Re13}.

However, the usual NC distributions written in terms of Gaussian densities lead to complexities in thermodynamic analysis since it requires us to work with incomplete gamma functions involving recurrent expressions, which must be treated in suitable asymptotic limits; as a consequence, it restricts the NC, breaks global quasi-homogeneity, and violates the Bekenstein-Hawking entropy relation \cite{Re14}, which are necessary to generate compelling descriptions of BH thermodynamics.

This strongly motivates the exploration of NC models that circumvent these issues while preserving regularity at the origin. Such a regular behavior can be partially achieved by introducing a Lorentzian density through an energy-momentum tensor that incorporates its NC effects. Unlike Gaussian densities that regularize metric functions, Lorentzian densities are mathematically simpler and lead to correction terms in the metric, providing a more natural framework for NC contexts.

As a result of considering NC spacetimes, the coordinates must satisfy an uncertainty principle. Thereby, it is not possible to define punctual locations in this case, as this would require measuring pairs of coordinates simultaneously, which is forbidden. Thus, usual NC descriptions involve distributions capable of removing essential singularities by avoiding punctual structures, since in GR singularities emerged from points. 
In this way, Lorentzian distributions replace the point-like source of the gravitational field to provide an alternative approach to studying NC effects in a manner that simplifies calculations in thermodynamics \cite{Re15}, since these distributions allow us to avoid some of the previously mentioned problems that appear in Gaussian distributions \cite{Re13}.

We write this work as follows. In Section 2, we present the noncommutative Reissner-Nordstr\"om (NCRN) metric given in a simple form using NC correction terms arising from a Lorentzian density in a first-order approach. We also obtain the conditions that follow from demanding that this BH is a quasi-homogeneous system, which are shown to be satisfied in general,  leading us to consider a quasi-homogeneous version of GTD. In Section 3, we write the basic ideas of GTD for quasi-homogeneous systems, which is our framework to analyze thermodynamics based on a pure geometric formalism capable of describing Legendre invariant configurations, implying that the thermodynamic potential can be chosen arbitrarily \cite{Re9}.
In Section 4, we calculate the thermodynamic properties of the NCRN solution. We focus on the phase transitions as described by the Ricci scalar divergences of the GTD metrics for the equilibrium space \cite{Re16}. In this sense, we obtain phase transition curves that reveal the role of the NC parameter $\Theta$.


\section{The Noncommutative Reissner-Nordstr\"{o}m Black Hole}

In this section, we present the main properties of the spacetime, which describes the gravitational field of a spherically symmetric mass distribution with constant electric charge, taking into account the effects of noncommutativity.

\subsection{The metric }

The NCRN spacetime is described by the spherically symmetric line element 
\begin{equation} \label{eq:1}
ds^2 = -f(r)dt^2 +
f^{-1}(r) dr^2 + r^2(d\theta^2 + sin^2 \theta d\phi^2), 
\end{equation}
where the metric function $f(r)$ is given by \cite{Re17}

\begin{align} \label{eq:2}
f(r) =\ & 1 - \frac{c}{r} + \frac{4\pi M\sqrt{\pi \Theta} - 2Q^2}{\pi^2 (\pi \Theta + r^2)} 
 \nonumber \\
& - \frac{2}{\pi^2 r} \left(2\pi M + \frac{Q^2}{\sqrt{\pi \Theta}} \right) \arctan\left(\frac{r}{\sqrt{\pi \Theta}}\right) 
 \nonumber \\
& + \frac{4Q^2}{\pi^2 r^2} \arctan^2 \left(\frac{r}{\sqrt{\pi \Theta}}\right),
\end{align}
where $M$ is the total mass and $Q$ the electric charge of the BH. 
Moreover, $c$ is an integration constant, and $\Theta$ the NC parameter generated by the Lorentzian distributions for mass and charge.

Notice that the exact metric function \eqref{eq:2} is derived from a phenomenological proposal among the NC models to describe non-punctual sources for the gravitational field in a stained spacetime. However, other more rigorous descriptions of NC spacetimes have been analyzed in  \cite{Re13}.

The explicit analysis of the above exact solution can   be performed only numerically due to the complexity of the metric function $f(r)$. To simplify the analysis, in this work, we will consider the limit $r\gg 1$ {with $c=-\frac{Q^2}{\sqrt{\pi^3 \Theta}}$} and expand the above metric function up to the third order in $1/r$. Then, we obtain 
\begin{equation} \label{eq:3}
f(r) = 1- \frac{2M}{r} + \frac{Q^2}{r^2} + \frac{8\sqrt{\Theta}M}{\sqrt{\pi}r^2} - \frac{4 \sqrt{\Theta}Q^2}{\sqrt{\pi}r^3},
\end{equation}
which can be interpreted as the NC generalization of the RN BH. 
In the limit of vanishing charge $Q=0$, from \eqref{eq:3} we obtain the NC Schwarzschild solution \cite{Re18}.
Notice that the solution \eqref{eq:3} can also be generalized to include a cosmological constant in a straightforward way \cite{Re17}. Nevertheless, in this work. we will not consider the cosmological constant as we are interested in understanding the influence of the NC parameter.

The solution  \eqref{eq:2} satisfies Einstein-Maxwell equations with an effective energy-momentum tensor, which can be represented as follows
\begin{align} \label{eq:4}
&  R_{\mu \nu} - 
\frac{1}{2}Rg_{\mu \nu} = 8 \pi \left(T^{(em)}_{\mu \nu} + {T_{\mu \nu}}^{(\Theta)}\right),
\\ & 
\frac{1}{\sqrt{-g}} \partial_{\mu}\left(\sqrt{-g}F^{\mu \nu}\right)=J^{\nu},
\end{align}
where 
the Faraday tensor $F^{\mu \nu}$ and the current density $J^\nu$ can be expressed as 
\begin{equation}  \label{eq:6}
    F^{\mu\nu}= E(r, \Theta) \begin{pmatrix}
0 & -1 & 0 & 0 \\
1 & 0 & 0 & 0 \\
0 & 0 & 0 & 0 \\
0 & 0 & 0 & 0
\end{pmatrix}
\end{equation}
\begin{equation} \label{eq:7}
    J^\nu = \rho_e (r,\Theta) \delta^\nu_0 ,
\end{equation}
respectively. Here, $E(r,\Theta)$ represents the electric field, which can be calculated from \eqref{eq:4} and takes the form
\begin{equation} \label{eq:8}
E(r,\Theta)=\frac{2Q}{\sqrt{\pi}r^2} \left[ \frac{1}{\sqrt{\pi}}\arctan \left(\frac{r}{\sqrt{\pi \Theta}}\right)-\frac{r \sqrt{\Theta}}{\pi \Theta + r^2} \right],
\end{equation}
with $E(r,\Theta)=\frac{Q}{r^2}$ when $r \rightarrow \infty$. 
Moreover, the density $\rho_e(r,\Theta)$ represents a Lorentzian charge distribution given by
\begin{equation} \label{eq:9}
\rho_e (r,\Theta)=\frac{4Q\sqrt{\Theta}}{\sqrt{\pi}(r^2+\pi \Theta)^2} .
\end{equation}
Then, the Maxwell tensor 
\begin{equation}  
 \label{eq:10}
T^{(em)}_{\mu\nu} \equiv \frac{1}{4 \pi} \left(g^{\rho \sigma} F_{\mu \rho} F_{\nu \sigma} - \frac{1}{4} g_{\mu \nu} F_{\rho \sigma}F^{\rho \sigma}\right)
\end{equation}
becomes
\begin{equation} \label{eq:11}
T^{\mu \ (em)}_{\ \nu} =\frac{E^2(r,\Theta)}{8\pi} diag(-1,-1,1,1).
\end{equation}
Finally, the effective energy-momentum tensor associated to the NC parameter can be expressed as 
\begin{equation} \label{eq:12}
T^{\mu\ (\Theta)}_{\ \nu} = diag \left(
-\rho^{(\Theta)},\rho^{(\Theta)},p_{\theta}^{(\Theta)},p_{\phi}^{(\Theta)}
\right).
\end{equation}
Interestingly, the NC contribution can be interpreted as a perfect fluid in terms of a Lorentzian mass density $\rho^{(\Theta)}$ and pressures $p^{(\Theta)}_{\theta}$ and $ p^{(\Theta)}_{\phi}$, which are written as
\begin{equation} \label{eq:13}
\rho^{(\Theta)} = \frac{M\sqrt{\Theta}}{\sqrt{\pi^3}(r^2+\pi \Theta)^2}  ,
\end{equation}
\begin{equation} \label{eq:14}
    p^{(\Theta)}_{\theta} = p^{(\Theta)}_{\phi},
\end{equation}
and satisfy the equation of state
\begin{equation} \label{eq:15}
    p^{(\Theta)}_{\theta} = - \rho^{(\Theta)} + \frac{r}{2} \partial_{r}\rho^{(\Theta)}.
\end{equation}

We see that the NC contribution can be interpreted as generating an effective density $\rho_e(r,\Theta)$, as expressed in Eq.(\ref{eq:9}), which gives rise to an effective current $J^\nu$ that modifies Maxwell's equations and induces the perfect-fluid-like tensor $T_{\mu\nu}^{(\Theta)}$.

Notice that in the limit 
$r \gg 1$, the electric field \eqref{eq:7} becomes 
\begin{equation} \label{eq:16}
   E(r,\Theta) \approx \frac{Q}{r^2} - \frac{4Q}{\sqrt{\pi}r^3}\sqrt{\Theta},
\end{equation}
and the components of the 
energy-momentum tensor reduce to 
\begin{equation} \label{eq:17}
    T^{\mu\ (em)}_{\ \nu} \approx \frac{Q^2}{8\pi r^4} \left(1-\frac{8\sqrt{\Theta}}{\sqrt{\pi }r}\right)diag(-1,-1,1,1)
\end{equation}
and 
\begin{equation} \label{eq:18}
    T^{\mu\ (\Theta)}_{\ \nu} \approx \frac{M\sqrt{\Theta}}{\sqrt{\pi^3} r^4}diag(-1,1,-3,-3). 
\end{equation}
This approximated energy-momentum tensor \eqref{eq:17} leads to the classical form of the Maxwell tensor when $\Theta=0$. Moreover, the components of the NC energy-momentum tensor satisfy the equations of state $p^{(\Theta)}_r = \rho^{(\Theta)}$ and $p^{(\Theta)}_\theta= -3 \rho^{(\Theta)}$, which can be interpreted as representing an exotic barotropic and anisotropic fluid, which behaves as stiff matter along the radial coordinate $r$ and as phantom energy along the angular directions. 

For the sake of completeness, we also quote the covariant components of the energy-momentum tensor: 
Using the limit $r \gg 1$ in $T^{(em)}_{\mu \nu}$ and $T^{(\Theta)}_{\mu \nu}$, we find:
\begin{equation} \label{eq:19}
T^{(em)}_{\mu \nu} \approx \frac{Q^2}{8\pi r^4} \left(1-\frac{8\sqrt{\Theta}}{\sqrt{\pi} r}\right)diag(1,1,r^2,r^2\sin^2 \theta),
\end{equation}
\begin{equation} \label{eq:20}
    T^{(\Theta)}_{\mu \nu} \approx \frac{M\sqrt{\Theta}}{\sqrt{\pi^3} r^4}diag(1,1,-3r^2,-3r^2 \sin^2 \theta).
\end{equation}
We point out that the addition of NC effects still allows us to use the usual structure for the Einstein-Hilbert action coupled to Maxwell electrodynamics $S=\int d^4x \sqrt{-g} (R + 2F_{\mu \nu}F^{\mu \nu})$ without modifications. Therefore, in this model, NC effects cannot be derived from the action but are contained in the energy-momentum tensor.

\subsection{Thermodynamics}

The Bekenstein-Hawking relation states that the entropy of a BH is proportional to the area of the event horizon $S=\frac{1}{4}A$, and represents the starting point of black hole thermodynamics. In the case of a spherically symmetric BH, we have $S= \pi r_H^2$, where $r_H$ is the horizon radius obtained from the condition $f(r_H)=0$ in \eqref{eq:3}.  The resulting equation can be rewritten as the mass function
\begin{equation} \label{eq:21}
M= \frac{\pi Q^2 \sqrt{S} + \sqrt{S^3} -4\pi Q^2 \sqrt{\Theta}}{2\sqrt{\pi S}(\sqrt{S}-4\sqrt{\Theta})},
\end{equation}
which is interpreted as the fundamental equation $M=M(S,Q,\Theta)$  for the NCRN BH.

In BH thermodynamics, rather than homogeneity, fundamental equations use to be quasi-homogeneity. Consequently,  \eqref{eq:21} should satisfy the quasi-homogeneous condition
\begin{equation} \label{eq:22}
M(\lambda^{\beta_a}E^a)=\lambda^{\beta_M}M(E^a),
\end{equation}
where $\lambda >0$ is a constant that rescales the variables $E^a$, while $\beta^a$ are real constants called quasi-homogeneous coefficients. It is easy to see that the fundamental equation \eqref{eq:21} satisfies the quasi-homogeneity condition only when the NC parameter $\Theta$ is taken into account as a thermodynamic variable. Therefore, the mass function \eqref{eq:21} satisfies the following condition
\begin{equation} \label{eq:23}
M(\lambda^{\beta_S}S, \lambda^{\beta_Q}Q, \lambda^{\beta_{\Theta}} \Theta)=\lambda^{\beta_M}M(S,Q,\Theta)
\end{equation}
if the quasi-homogeneity  coefficients are related by
\begin{equation} \label{eq:24}
    \beta_S =\beta_{\Theta}, \ \beta_S= 2\beta_{Q}, \ \beta_S=2\beta_{M}.
\end{equation}

Quasi-homogeneity is a property of thermodynamic systems, which does not affect the laws of thermodynamics. Nevertheless, it does modify the usual thermodynamic identities, such as the Euler identity \cite{Re16}
\begin{equation} \label{eq:25}
\sum_{a} \beta_a E^a \frac{\partial M}{\partial E^a} = \beta_M M 
\end{equation}
and the Gibbs-Duhem identity
\begin{equation} \label{eq:26}
    \sum_a (\beta_a-\beta_M)\frac{\partial M}{\partial E^a} dE^a + \sum_{a,b}\beta_a E^a \frac{\partial^2M}{\partial E^a\partial E^b}dE^b = 0.
\end{equation}
When $\beta_a = \beta_M=1$ the homogeneous case is recovered. 
Thus, in the NCRN BH, the Euler identity in the mass representation is given by:
\begin{equation} \label{eq:27}
2S \frac{\partial M}{\partial S} + Q \frac{\partial M}{\partial Q} +2\Theta \frac{\partial M}{\partial \Theta}=M .
\end{equation}

\section{Quasi-Homogeneous Geometrothermodynamics in brief}

The GTD formalism aims to describe the physical properties of thermodynamic systems using a purely geometrical and Legendre invariant description in terms of concepts of differential geometry. Here, it is convenient to represent Legendre
transformations as coordinate transformations that leave the geometric structure of a differential manifold invariant \cite{Re9}. It is achieved by 
defining a phase space $\mathcal{T}$, which is equipped with Riemannian metrics that are invariant under the action of Legendre transformations, i.e., their geometric properties do not depend on the choice of thermodynamic potential. The phase space contains a subspace called the equilibrium space $\mathcal{E}$ whose points correspond to the equilibrium states of the system.

To define the phase space $\mathcal{T}$ in GTD, we consider a system with $n$ thermodynamic degrees of freedom. As coordinates in $\mathcal{T}$, we introduce the thermodynamic potential $\Phi$, the extensive variables $E^a$,
and the intensive variables $I^a$, where $a=1,...,n$, which implies having a $(2n+1)$-dimensional phase space with coordinates $Z^A= \{\Phi,E^a,I_a\}$. The
geometric properties of $\mathcal{T}$ are invariant under the action of general diffeomorphisms of the
form $Z^A \rightarrow Z^{A'}=Z^{A'}(Z^A)$ with 
\begin{equation} \label{eq:28}
\left| \frac{\partial Z^{A'}}{\partial Z^{A}} \right | \neq 0 .
\end{equation}

In particular, Legendre transformations can be represented as coordinate transformations in the form \cite{Re19}, \cite{Re20}
\begin{equation} \label{eq:29}
    \{Z^A\} \to \{\tilde{Z}^A\} = \{\tilde{\Phi}, \tilde{E}^a, \tilde{I}_a\}
\end{equation}
with
\begin{equation} \label{eq:30}
    \Phi = \tilde{\Phi} - \tilde{E}^k \tilde{I}_k, \quad E^i = -\tilde{I}_i, \quad E^j = \tilde{E}^j, \quad I_i = \tilde{E}^i, \quad I_j = \tilde{I}^j
\end{equation}

where $i \cup j$ is any disjoint decomposition of the set of indices $1, \dots, n$ and $k, l = 1, \dots, i$. It follows that for $i = 1, \dots, n$ and $i = \emptyset$, we obtain the total Legendre transformation and the identity, respectively.

We observe that the above Legendre transformation  can exchange extensive and intensive variables, leading to a loss of their physical meaning, which can be seen as an analogy to the canonical transformations in classical mechanics.

To ensure Legendre invariance of the differential manifold $\mathcal{T}$, we endow it with a Riemannian metric $G^{AB}$ whose  functional dependence must remain unchanged under the action Legendre coordinate transformations \eqref{eq:29} and \eqref{eq:30}. By applying the Legendre transformation to an arbitrary $G^{AB}$, the conditions for its invariance  turn out to be satisfied by the following line elements \cite{Re16}, \cite{Re21}
\begin{equation} \label{eq:31}
    G^{\mathrm{I}} = (\mathrm{d}\Phi - I_a \mathrm{d}E^a)^2 + (\zeta_{ab} E^a I^b) (\delta_{cd} \mathrm{d}E^c \mathrm{d}I^d),
\end{equation}
\begin{equation} \label{eq:32}
    G^{\mathrm{II}} = (\mathrm{d}\Phi - I_a \mathrm{d}E^a)^2 + (\zeta_{ab} E^a I^b) (\eta_{cd} \mathrm{d}E^c \mathrm{d}I^d),
\end{equation}
\begin{equation} \label{eq:33} 
G^{\mathrm{III}} = (\mathrm{d}\Phi - I_a \mathrm{d}E^a)^2 + \sum_{a=1}^{n} (\zeta_a (E^a I^a)^{2k+1} \mathrm{d}E^a \mathrm{d}I^a)
\end{equation}

where $\delta_{ab} = \mathrm{diag}(1, \dots, 1)$, $I^a = \delta^{ab} I_b$, $\eta_{ab} = \mathrm{diag}(-1, 1, \dots, 1)$, $\zeta_a$ are real constants, $\zeta_{ab}$ is a diagonal $n \times n$ real matrix, and $k$ is an integer.

From  Darboux's theorem \cite{Re22}, the odd-dimensional differential manifold $\mathcal{T}$ can be endowed in a canonical way with a contact structure determined by the 1-form:
\begin{equation} \label{eq:34}
    \Theta_{\mathcal{T}} = \mathrm{d}\Phi - I_a \mathrm{d}E^a, \quad \text{with} \quad \Theta_{\mathcal{T}} \wedge (\mathrm{d}\Theta_{\mathcal{T}})^{\wedge n} \neq 0
\end{equation}
which is also Legendre invariant, since under the action of the Legendre transformation $Z^A \to \tilde{Z}^A$, its functional dependence remains unchanged as:
\begin{equation} \label{eq:35} 
    \Theta_{\mathcal{T}} \to \hat{\Theta}_{\mathcal{T}} = \mathrm{d}\tilde{\Phi} - \tilde{I}_a \mathrm{d}\tilde{E}^a.
\end{equation}
$\Theta_{\mathcal{T}}$ is a canonical contact 1-form that is a significant ingredient in GTD since it represents the first law of thermodynamics when projected on the equilibrium space $\mathcal{E}$.
Moreover, $\mathcal{E}$ is an $n$-dimensional subspace of the phase space $\mathcal{T}$ defined by the smooth embedding map
\begin{equation} \label{eq:36}
    \varphi: \mathcal{E} \to \mathcal{T}, \quad \text{i.e.}, \quad Z^A = \{\Phi(E^a), E^a, I_a(E^b)\}
\end{equation}
which satisfies the condition
\begin{equation} \label{eq:37}
    \varphi^*(\Theta_{\mathcal{T}}) = 0, \quad \text{i.e.}, \quad \mathrm{d}\Phi - I_a \mathrm{d}E^a = 0,
\end{equation}
where $\varphi^*$ is the pullback of $\varphi$. Since the explicit form of the embedding map requires that the thermodynamic potential be a function of the variables $E^a$, we obtain the following relationship
\begin{equation} \label{eq:38}
    \mathrm{d}\Phi = \frac{\partial\Phi}{\partial E^a} \mathrm{d}E^a, \quad \text{i.e.}, \quad I_a = \frac{\partial\Phi}{\partial E^a}
\end{equation}
The embedding map $\varphi$ induces the equation $\Phi = \Phi(E^a)$, which is the fundamental equation. Consequently, the relationships \eqref{eq:38} in GTD represent the first law of thermodynamics and the equilibrium conditions of the system.

On the other hand, the metric of $\mathcal{E}$ is also induced canonically by the embedding map $\varphi$ using the corresponding pullback $\varphi^*$ as follows:
\begin{equation} \label{eq:39}
    g = g_{ab} \mathrm{d}E^a \mathrm{d}E^b = \varphi^*(G) = \varphi^* (G_{AB} \mathrm{d}Z^A \mathrm{d}Z^B)
\end{equation}
whose components are
\begin{equation}  \label{eq:40}
    g_{ab} = G_{AB} \frac{\partial Z^A}{\partial E^a} \frac{\partial Z^B}{\partial E^b}.
\end{equation}
As a consequence, from Eqs.\eqref{eq:31}-\eqref{eq:33}, we obtain the following independent and invariant metrics for the equilibrium space $\mathcal{E}$ which are the final result of GTD:
\begin{equation} \label{eq:41}
    g^{\mathrm{I}} = \sum_{a,b,c=1}^{n} \left(\beta_c E^c \frac{\partial \Phi}{\partial E^c} \right) \frac{\partial^2 \Phi}{\partial E^a \partial E^b} \mathrm{d}E^a \mathrm{d}E^b,
\end{equation}
\begin{equation} \label{eq:42} 
    g^{\mathrm{II}} = \sum_{a,b,c,d=1}^{n} \left(\beta_c E^c \frac{\partial \Phi}{\partial E^c} \right) \eta^d_a \frac{\partial^2 \Phi}{\partial E^b \partial E^d} \mathrm{d}E^a \mathrm{d}E^b,
\end{equation}
\begin{equation} \label{eq:43}
    g^{\mathrm{III}} = \sum_{a,b=1}^{n} \left(\beta_a E^a \frac{\partial \Phi}{\partial E^a} \right) \frac{\partial^2 \Phi}{\partial E^a \partial E^b} \mathrm{d}E^a \mathrm{d}E^b.
\end{equation}
Here, $\eta^a_c = \mathrm{diag}(-1, 1, \dots, 1)$.

The free parameters of the line elements $G^{\mathrm{I}}$, $G^{\mathrm{II}}$, and $G^{\mathrm{III}}$ have been chosen in terms of the quasi-homogeneous coefficients as
\begin{equation} \label{eq:44}
    \zeta_a = \beta_a, \quad \zeta_{ab} = \mathrm{diag}(\beta_1, \dots, \beta_n), \quad k = 0
\end{equation}
This choice results from the requirement that all three GTD metrics be applied simultaneously to the same thermodynamic system and yield compatible results. It shows that to have a comprehensive GTD analysis, it is necessary to know an explicit form of the fundamental equation $\Phi = \Phi(E^a)$, which allows to describe all the thermodynamic properties of the system.

\section{Phase transitions in NCRN BHs}

Following the results of classical equilibrium thermodynamics, 
Davies proposed using the heat capacity divergences to determine the phase transitions of BHs \cite{Re5}. On the basis of a pure thermodynamic analogy, phase transitions are often interpreted
as transitions from cold to hot BHs, as well as from small to large black BHs. 
However, the inner microstructure of BHs in terms of statistical mechanics remains a mystery, so a complete description
of BH phase transitions is still unavailable, but it is possible to explore our understanding of them using novel formalisms to address thermodynamics; for example, GTD has shown to reveal second-order phase transitions \cite{Re23} that are not predicted by heat capacity  \cite{Re10}.

To calculate the three invariant metrics of GTD \eqref{eq:41}-\eqref{eq:43}, we choose the mass $M$ as the thermodynamic
potential $\Phi$ and $E^a = (S,Q,\Theta )$ as the extensive thermodynamic variables. Consequently, we have:
\begin{align} \label{eq:45}
g^{I}= \ &  \left(\beta_S S \frac{\partial M}{\partial S } + \beta_Q Q \frac{\partial M}{\partial Q } + \beta_\Theta \Theta \frac{\partial M}{\partial \Theta }\right)
\nonumber \\ &
\times \bigg[ \frac{\partial^2M}{\partial S^2} dS^2 + \frac{\partial^2M}{\partial Q^2} dQ^2 + \frac{\partial^2M}{\partial \Theta^2} d\Theta^2
\nonumber \\ &
+ 2\bigg( \frac{\partial^2M}{\partial S \partial Q} dSdQ +  \frac{\partial^2M}{\partial S \partial \Theta} dSd\Theta + \frac{\partial^2M}{\partial Q \partial \Theta} dQd\Theta\bigg)\bigg ],
\end{align}
\begin{align} \label{eq:46}
g^{II}= \ &  \bigg(\beta_S S \frac{\partial M}{\partial S } + \beta_Q Q \frac{\partial M}{\partial Q } + \beta_\Theta \Theta \frac{\partial M}{\partial \Theta }\bigg)
\nonumber \\ &
\times \bigg[ -\frac{\partial^2M}{\partial S^2} dS^2 + \frac{\partial^2M}{\partial Q^2} dQ^2 + \frac{\partial^2M}{\partial \Theta^2} d\Theta^2 
\nonumber \\ &
+ 2\frac{\partial^2M}{\partial Q \partial \Theta} dQd\Theta\bigg],
\end{align}
\begin{align} \label{eq:47}
g^{III}= \ & \beta_S S \frac{\partial M}{\partial S }\frac{\partial^2M}{\partial S^2} dS^2 + \beta_Q Q \frac{\partial M}{\partial Q } \frac{\partial^2M}{\partial Q^2} dQ^2 
\nonumber \\ &
+\beta_\Theta \Theta \frac{\partial M}{\partial \Theta } \frac{\partial^2M}{\partial \Theta^2} d\Theta^2 \nonumber \\ & 
+ \bigg(\beta_S S \frac{\partial M}{\partial S } + \beta_\Theta \Theta \frac{\partial M}{\partial \Theta }\bigg)
\frac{\partial^2M}{\partial S \partial \Theta} dS d\Theta \nonumber \\ &
+ \bigg(\beta_Q Q \frac{\partial M}{\partial Q } + \beta_\Theta \Theta \frac{\partial M}{\partial \Theta }\bigg)
\frac{\partial^2M}{\partial Q \partial \Theta} dQ d\Theta 
\nonumber  \\  &
+ \bigg(\beta_S S \frac{\partial M}{\partial S } + \beta_Q Q \frac{\partial M}{\partial Q }\bigg)
\frac{\partial^2M}{\partial S \partial Q} dS dQ
\end{align}
These metrics replicate the $2-dim$ result found in the NC Schwarzschild BH when $Q=0$ \cite{Re18}. 

Furthermore, these metrics can be written explicitly using the fundamental equation \eqref{eq:21} and the Euler identities \eqref{eq:25} and \eqref{eq:27}. We obtain
\begin{align} \label{eq:48}
g^{I}= \ & \left( \frac{\beta_M \left(\sqrt{S^3} + \pi Q^2 (\sqrt{S} - 4\sqrt{\Theta})\right)}{16 \pi \sqrt{S^9}(\sqrt{S}-4\sqrt{\Theta})^4 } \right)
\nonumber \\ &
\times [(-S^4+3\pi S^3Q^2+12\sqrt{S^7 \Theta}  
\nonumber \\ & 
- 36\pi \sqrt{S^5 \Theta}Q^2 +144 \pi S^2Q^2 \Theta 
\nonumber \\ & - 192\pi \sqrt{S^3 \Theta^3}Q^2 ) dS^2
\nonumber \\ &
+8\pi \sqrt{S^7}(\sqrt{S}-4\sqrt{\Theta})^3 dQ^2
\nonumber \\ &
- \frac{4S^5}{\sqrt{\Theta^3}} (\sqrt{S}-12\sqrt{\Theta})d\Theta^2 
- 64S^4 dSd\Theta
\nonumber \\ &
-8\pi \sqrt{S^5}Q(\sqrt{S}-4\sqrt{\Theta})^3 dSdQ]
\end{align}
\begin{align} \label{eq:49}
g^{II}= \ & \left( \frac{\beta_M \left( \sqrt{S^3} + \pi Q^2 (\sqrt{S} - 4\sqrt{\Theta}) \right)}{16 \pi \sqrt{S^9}(\sqrt{S}-4\sqrt{\Theta})^4 } \right)
\nonumber \\ &
\times [(S^4 -3\pi S^3Q^2 -12\sqrt{S^7 \Theta} 
\nonumber \\ & 
+ 36\pi \sqrt{S^5 \Theta} Q^2
-144 \pi S^2Q^2 \Theta 
\nonumber \\ &
+ 192\pi \sqrt{S^3 \Theta^3}Q^2) dS^2
\nonumber \\ &
+8\pi \sqrt{S^7} (\sqrt{S}-4\sqrt{\Theta})^3 dQ^2
\nonumber \\ &
- \frac{4S^5}{\sqrt{\Theta^3}} (\sqrt{S}-12\sqrt{\Theta})d\Theta^2]
\end{align}
\begin{align} \label{eq:50}
g^{III}= \ & \left( \frac{\beta_M}{16 \pi S^3 (\sqrt{S}-4\sqrt{\Theta})^5} \right)
\times [(3\pi^2 Q^4 (\sqrt{S}-4\sqrt{\Theta})^5
\nonumber \\ &
-4\pi \sqrt{S^3} Q^2 (\sqrt{S}-4\sqrt{\Theta})^2 (S-12\sqrt{S \Theta}+24\Theta) 
\nonumber \\ &
+ \sqrt{S^7}(S-20\sqrt{S \Theta} + 96\Theta))dS^2 
\nonumber \\ &
- 16 \pi^2 S^2 Q^2 (\sqrt{S}-4\sqrt{\Theta})^5 dQ^2
\nonumber \\ &
-\frac{16S^5}{\Theta}(\sqrt{S}-12\sqrt{\Theta})d\Theta^2
\nonumber \\ &
+ 4\pi Q(\sqrt{S^5} (\sqrt{S}-8\sqrt{\Theta})(\sqrt{S}-4\sqrt{\Theta})^3 
\nonumber \\ &
+ \pi SQ^2 (\sqrt{S}-4\sqrt{\Theta})^5)dSdQ 
\nonumber \\ &
-32 S (\sqrt{S}-4 \sqrt{\Theta})(S^3 -\pi S^2 Q^2
\nonumber \\ &
+ 4\pi \sqrt{S^3 \Theta}Q^2)dS d\Theta]
\end{align}
  
Here, beyond trivial singularities, all these metrics contain the same coordinate singularity $S=16\Theta$ inherited by the singularity in the fundamental equation \eqref{eq:21}.

From the GTD metrics \eqref{eq:45}-\eqref{eq:47} we can calculate the Ricci scalar singularities, which are described by the following conditions:
  \begin{align} \label{eq:51}
I: \frac{\partial^2M}{\partial Q^2} \left(\frac{\partial^2M}{\partial S \partial \Theta}\right)^2 - 2\frac{\partial^2M}{\partial Q \partial \Theta}\frac{\partial^2M}{\partial S \partial \Theta}\frac{\partial^2M}{\partial S \partial Q}  \nonumber \\ + \frac{\partial^2M}{\partial \Theta^2}\left(\frac{\partial^2M}{\partial S \partial Q}\right)^2 
\nonumber \\ 
+\frac{\partial^2M}{\partial S^2}  \left(\left(\frac{\partial^2M}{\partial Q \partial \Theta}\right)^2
-\frac{\partial^2M}{\partial Q^2}\frac{\partial^2M}{\partial \Theta^2}\right) &= 0 , 
\end{align}
\begin{align} \label{eq:52}
II: \frac{\partial^2M}{\partial S^2} \left( \left(\frac{\partial^2M}{\partial Q \partial \Theta}\right)^2 -  \frac{\partial^2M}{\partial Q^2 } \frac{\partial^2M}{\partial \Theta^2}  \right) = 0  ,
\end{align}
\begin{align} \label{eq:53}
III: \frac{\partial^2M}{\partial S^2} = \frac{\partial^2M}{\partial S \partial Q} = \frac{\partial^2M}{\partial S \partial \Theta} = 0
\nonumber \\ & \text{or} \nonumber \\
\frac{\partial^2M}{\partial S \partial Q}=\frac{\partial^2M}{\partial Q^2}=0
\nonumber \\ & \text{or} \nonumber \\
\frac{\partial^2M}{\partial S \partial \Theta}=\frac{\partial^2M}{\partial \Theta^2}=0
\nonumber \\ & \text{or} \nonumber \\
\frac{\partial^2M}{\partial Q^2}=\frac{\partial^2M}{\partial \Theta^2}=0
\end{align}
The first condition \eqref{eq:51} corresponds to the zeros of the determinant of the Hessian matrix of the mass, $det [Hess (M(S,Q,\Theta)) ]=0$, which leads to a stability condition for thermodynamical systems with three degrees of freedom. This can be viewed as a generalization of the 2D system \cite{Re18}. The second condition \eqref{eq:52} recovers the expected phase transition $M_{,SS}=0$ inspired by the divergence of the heat capacity \cite{Re5}. The third condition \eqref{eq:53} is obtained by including the other conditions in a similar way as shown in the three degree of freedom analysis \cite{Re24}, but incorporating an NC parameter. 
  
Therefore, using the fundamental equation \eqref{eq:21} in \eqref{eq:51}-\eqref{eq:53}, we find that the second and third conditions are equivalent so that only the first two conditions lead to non-trivial phase transitions. These conditions can be easily written in terms of the electric charge $Q$ as:
\begin{align} \label{eq:54}
I: Q^{(I)}=\pm S^{3/2}\sqrt{\frac{\sqrt{S}-16\sqrt{\Theta}}{\pi (\sqrt{S}-4\sqrt{\Theta})(\sqrt{S}-12\sqrt{\Theta})}},
\end{align}
\begin{align} \label{eq:55}
II\equiv III :Q^{(II)}=\pm S\sqrt{\frac{ \sqrt{S}-12\sqrt{\Theta} }{3\pi (\sqrt{S}-4\sqrt{\Theta})^3}}.
\end{align}
Thus, the two conditions \eqref{eq:54}, \eqref{eq:55} represent the values of the electric charge for which the GTD phase transitions occur in a system described by the NCRN BH.
It is straightforward to verify that the NC Schwarzschild phase transitions $S=144\Theta, 256\Theta$ \cite{Re18} can be completely recovered using $Q=0$ in \eqref{eq:54} and \eqref{eq:55}. 

We illustrate the behavior of the functions  \eqref{eq:54} and  \eqref{eq:55} in Figures 1 and 2, respectively.  Unlike the uncharged scenario \cite{Re18}, here the phase transition conditions are determined by a collection of points for the charge $Q$. In the commutative case $\Theta=0$, one obtains the values of the classical RNBH phase transition $Q^{(I)}_{RN}= \pm \sqrt{\frac{S}{\pi}}$ and $Q^{(II)}_{RN}= \pm \frac{Q^{(I)}_{RN}}{\sqrt{3}}$, which can be associated with the extremal case and the divergence in the heat capacity of the classical RNBH \cite{Re25}, respectively. 
Furthermore, in the NC scenario, phase transitions occur only for NC values of $\Theta$ in which the charge function $Q (S, \Theta)$ is well-defined. In the particular case $S=1$, 
these values are  $0 \leq \Theta \leq \frac{1}{256}, \frac{1}{144} < \Theta < \frac{1}{16}$ for \eqref{eq:54} and  $0 \leq \Theta \leq \frac{1}{144}, \Theta > \frac{1}{16}$ for \eqref{eq:55}.

\begin{figure}[h!]
    \centering
    \includegraphics[width=0.4\textwidth]{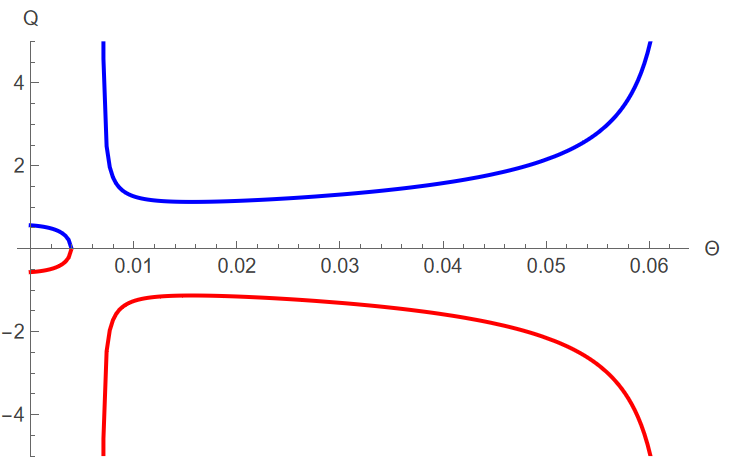}
    \caption{Electric charge $+Q^{I}$ (blue) and $-Q^{I}$ (red) represented as functions by parts showing the values where phase transitions can occur in a NCRN BH, according to the condition \eqref{eq:54}. Here, $S=1$.}
    \label{fig:miprimeraimagen}
\end{figure}

\begin{figure}[h!]
    \centering
    \includegraphics[width=0.4\textwidth]{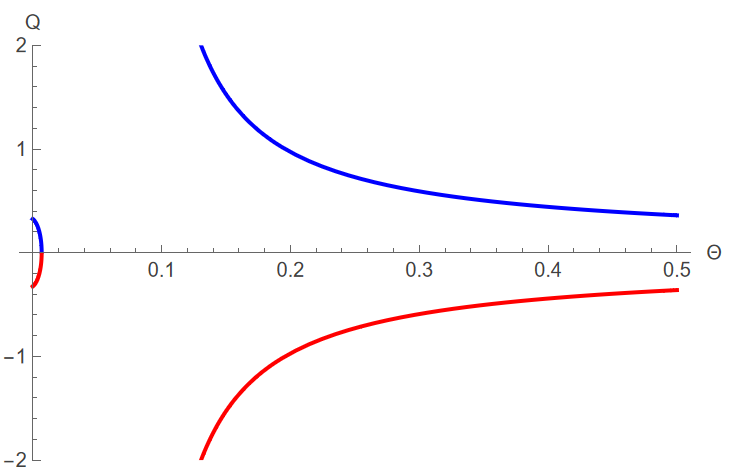}
    \caption{Analogously to Figure 1, this image represents the electric charge $+Q^{II}$ (blue) and $-Q^{II}$ (red) for the possible GTD phase transitions according to the condition \eqref{eq:55}.}
    \label{fig:misegundaimagen}
\end{figure}

To investigate the physical meaning of the curvature singularities,  we consider the first law of BH thermodynamics, which can be obtained from the general expression of GTD according to \eqref{eq:37} using the mass representation:
\begin{equation} \label{eq:56}
dM=TdS+\phi dQ +Nd\Theta
\end{equation}
Here, $T$ is the Hawking temperature of the BH, $\phi$ is the electric potential, and $N$ is the NC thermodynamic variable dual to $\Theta$. Using the mass function \eqref{eq:21}, we obtain 
\begin{align} \label{eq:57}
\nonumber \\& T= \frac{\partial M}{\partial S} 
\nonumber \\ & 
= \frac{\sqrt{S}(\sqrt{S}-8\sqrt{\Theta})(S-\pi Q^2)-16\pi Q^2 \Theta}{4 \sqrt{\pi S^3}(\sqrt{S}-4\sqrt{\Theta})^2}
\end{align}
\begin{equation} \label{eq:58}
\phi=\frac{\partial M}{\partial Q}=\frac{\sqrt{\pi} Q}{\sqrt{S}}
\end{equation}
\begin{equation} \label{eq:59}
N=\frac{\partial M}{\partial \Theta}=\frac{S }{\sqrt{\pi  \Theta}(\sqrt{S}-4\sqrt{\Theta})^2}
\end{equation}
The Hawking temperature must satisfy $T>0$ to be a physical quantity, which is satisfied by the following conditions considering $\Theta \geq 0$:
\begin{align} \label{eq:60}
& S > 64\Theta,
\nonumber \\ &
\left|Q\right| < \sqrt{\frac{S^2(S-48\Theta)-128\sqrt{(S\Theta)^3}}{\pi(S-16\Theta)^2}}
\end{align}

The three quantities $T, \phi$, and $N$ are  intensive variables within BH thermodynamics. We note that $T$ and $N$ show a divergence for the value $S=16\Theta$, but this is a non-physical divergence since it would require an infinite mass in the fundamental equation \eqref{eq:21} employed for this calculation.

\begin{figure}[h!]
    \centering
    \includegraphics[width=0.5\textwidth]{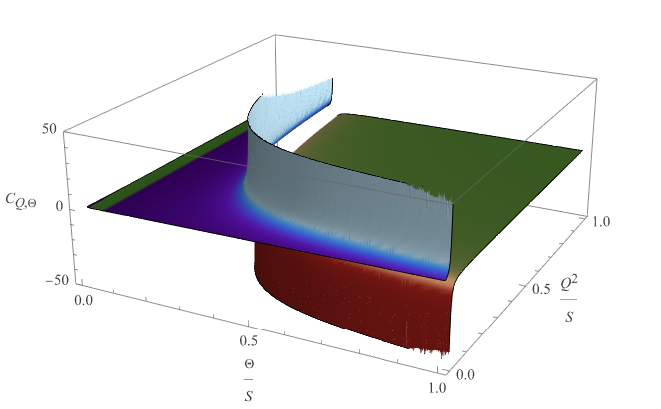}
    \caption{The heat capacity $C_{Q,\Theta}=C_{Q,\Theta} (\frac{\Theta}{S},\frac{Q^2}{S})$ of the NCRN BH. Here, $C_{Q,\Theta} < 0$ (red) indicates unstable configurations, while the stable values appear for $C_{Q,\Theta} > 0$ (blue). The phase transition curve \eqref{eq:62} can be visualized for the positive values of the axis since $\frac{Q^2}{S}> 0$.}
    \label{fig:miterceraimagen}
\end{figure}

To unravel the nature of phase transitions, we consider the heat capacity $C_{Q,\Theta}$ at constant values of the electric charge $Q$ and the NC parameter $\Theta$:
\begin{align} \label{eq:61}
& C_{Q, \Theta} \equiv T \left( \frac{\partial S}{\partial T} \right) \equiv \frac{\frac{\partial M}{\partial S}}{\left( \frac{\partial^2 M}{\partial S^2} \right)} 
\nonumber \\ &
= \frac {2S(4\sqrt{\Theta}-\sqrt{S})\left(  \sqrt{S}(\sqrt{S}-8\sqrt{\Theta})(S-\pi Q^2) -16\pi Q^2 \Theta \right)}{S(12\sqrt{\Theta}-\sqrt{S})(S-3\pi Q^2)- 48\pi Q^2 \Theta (-3\sqrt{S}+4\sqrt{\Theta})} 
\end{align}
Clearly, phase transitions occur when the denominator of \eqref{eq:61} diverges, which corresponds to $M_{,SS}=0$. This condition leads directly to the function \eqref{eq:55} for the electric charge $Q$. This shows that the curvature singularity (\ref{eq:55}) corresponds to a second-order phase transition. 
The curve that gives rise to divergences in $C_{Q,\Theta}$ is   
\begin{equation}\label{eq:62}
\frac{Q^2}{S}=
\frac{12\sqrt{\frac{\Theta}{S}}-1}{3\pi \left(-1+12\sqrt{\frac{\Theta}{S}}-48 \frac{\Theta}{S} + 64 {(\frac{\Theta}{S}})^{3/2}\right)},
\end{equation}
which, for $\Theta=0$, reproduces the aforementioned phase transition with charge $Q=\sqrt{\frac{S}{3\pi}}$ of the classical RNBH. Likewise, for $Q=0$, it yields the NC Schwarzschild phase transition at $S=144 \Theta$. We show the behavior of the heat capacity $C_{Q,\Theta}$ in Fig. \ref{fig:miterceraimagen}. 

\section{Conclusions}

Within the GTD formalism, the NC effects on the thermodynamics of the NCRN BH were examined, focusing on those properties associated with phase transitions.

Here, we use a metric function in the large-radius regime $r \gg 1$, which allows us to use the mass $M$ as the fundamental equation, without further approximations in $\sqrt{\Theta}$. This mass contains all the thermodynamic properties of the system and turns out to be a quasi-homogeneous function. As a consequence, the NC parameter $\Theta$ must be treated as an independent thermodynamic variable, playing a significant role that influences all the properties of the BH.

To construct this analysis, we start from the GTD framework, which provides an elegant approach to studying thermodynamics from a purely geometric and Legendre-invariant perspective. Using the curvature singularities of the three invariant GTD metrics, we identify the phase transitions of the NCRN BH.

Having obtained the fundamental equation in the mass representation, we write the three
metrics of the equilibrium space and derive the corresponding Ricci scalar singularities, which are physically interpreted as the stability and  phase transition conditions. We find two
distinct non-trivial phase transition conditions, which can be conveniently written in terms of electric charge functions $Q(S,\Theta)$, yielding 3D curves that indicate the regions where phase transitions may occur. These results are consistent with the classical RN BH case $\Theta=0$, where we recover the charges associated with the extremal configurations $\left| Q \right| =M$ as well as the singularities of the classical heat capacity $C_Q$. Meanwhile, in the uncharged case $Q=0$, as expected, we recover singular points previously found in the NC Schwarzschild analysis $S=144\Theta, 256 \Theta$.
In addition, we calculate other thermodynamic properties, such as temperature and heat capacity, to establish whether these conditions are reproduced by the GTD formalism. 

All of these results exhibit that, given a fundamental equation, GTD can be employed to analyze exotic thermodynamic systems such as BHs with multiple parameters. An interesting extension for future work consists of examining higher-dimensional NC BHs, such as those coupled to non-linear electrodynamics, and deriving their thermodynamic properties using the quasi-homogeneous GTD statements.

\section*{Acknowledgments}
This work was partially supported by UNAM-DGAPA-PAPIIT, grant No. 108225, and Conahcyt, grant No. CBF-2025-I-243.

\end{document}